\documentclass[12pt]{article}
\usepackage{amsmath,amsthm,amssymb,amsfonts}
\usepackage{bm}
\usepackage{mathtools}
\usepackage{mathrsfs}
\usepackage{graphicx}
\usepackage{algorithm}
\usepackage{algorithmic}
\usepackage{natbib}

\usepackage{hyperref}
\hypersetup{
  colorlinks=true,
  linkcolor=black,
  citecolor=blue,
}
\usepackage{xcolor}
\usepackage{caption}
\usepackage{booktabs}
\usepackage{textcomp}

\usepackage[scaled,mono=false]{libertine}
\usepackage{libertinust1math}
\usepackage{microtype}

\newcommand{\blind}{0}
\newtheorem{theorem}{Theorem}[section]

\newtheorem{definition}{Definition}[section]

\newtheorem{lemma}{Lemma}[section]

\newcommand{\bbeta}{\boldsymbol{\beta}}
\newcommand{\bgamma}{\boldsymbol{\gamma}}

\newcommand{\btheta}{\boldsymbol{\theta}}
\newcommand{\bxi}{\boldsymbol{\xi}}

\newcommand{\boldeta}{\boldsymbol{\eta}}

\newcommand{\bLambda}{\boldsymbol{\Lambda}}

\newcommand{\bA}{\mathbf{A}}
\newcommand{\bI}{\mathbf{I}}

\newcommand{\bX}{\mathbf{X}}

\newcommand{\bZ}{\mathbf{Z}}

\newcommand{\ba}{\mathbf{a}}

\newcommand{\bc}{\mathbf{c}}
\newcommand{\bd}{\mathbf{d}}

\newcommand{\bu}{\mathbf{u}}
\newcommand{\bv}{\mathbf{v}}
\newcommand{\bw}{\mathbf{w}}
\newcommand{\bx}{\mathbf{x}}
\newcommand{\by}{\mathbf{y}}
\newcommand{\bz}{\mathbf{z}}

\newcommand{\real}{\mathbb{R}}

\newcommand{\T}{\scriptscriptstyle{\text{T}}}

\DeclareMathOperator*{\argmin}{arg\,min}

\usepackage{letltxmacro}
\makeatletter
\let\oldr@@t\r@@t
\def\r@@t#1#2{%
\setbox0=\hbox{$\oldr@@t#1{#2\,}$}\dimen0=\ht0
\advance\dimen0-0.2\ht0
\setbox2=\hbox{\vrule height\ht0 depth -\dimen0}%
{\box0\lower0.4pt\box2}}
\LetLtxMacro{\oldsqrt}{\sqrt}
\renewcommand*{\sqrt}[2][\ ]{\oldsqrt[#1]{#2}}
\makeatother

\begin{document}

\def\spacingset#1{\renewcommand{\baselinestretch}%
  {#1}\small\normalsize}%
\spacingset{1}


\if0\blind {%
  \title{Adaptive Sparse Group Lasso Penalized Quantile Regression via Dual ADMM}%
  \author{Huayan Kou, Yuwen Gu, Yi Lian, Rui Zhang and Jun Fan \\ \\
    Department of Applied Statistics, Hebei University of Technology \\
    \small{E-mail: 202431103014@stu.hebut.edu.cn}\\
    Department of Statistics, University of Connecticut\\
    \small{E-mail: yuwen.gu@uconn.edu}\\
    Department of Biostatistics, Epidemiology and Informatics, University of Pennsylvania\\
    \small{E-mail: yi.lian@pennmedicine.upenn.edu}\\
    Institute of Mathematics, Hebei University of Technology\\
    \small{E-mail: ruizhang199@163.com}\\
    Institute of Mathematics, Hebei University of Technology\\
    \small{E-mail: fanjunmath@hotmail.com}\\
  }%
  \date{}%
  \maketitle%
}%
\fi


\bigskip
\begin{abstract}
  Sparse penalized quantile regression provides an effective framework for variable
  selection and robust estimation in high-dimensional data analysis. When explanatory variables are organized into groups, achieving sparsity both within and between groups is essential. However, existing quantile regression methods often fail to meet this dual objective. To address this gap, we consider the adaptive sparse group lasso penalized quantile regression, which integrates adaptive lasso and adaptive group lasso penalties. We optimize the model parameters via the alternating direction method of
  multipliers (ADMM) applied to the dual problem, and establish global convergence. Through extensive simulation studies and real data analyses, we demonstrate (i) the efficacy of the resulting estimator in achieving simultaneous within- and between-group sparsity, and (ii) the computational efficiency of our algorithm relative to existing alternatives.
\end{abstract}

\noindent%

\vfill

\spacingset{1.25} 
\clearpage

\section{Introduction}
Quantile regression, pioneered by \cite{koenker1978regression}, has
become a widely used tool in statistical analysis. Its appeal lies in
its robustness and its ability to reveal aspects of the conditional
distribution of the response given the covariates that are
inaccessible through least squares regression. Consider the linear quantile regression model
\begin{equation}\label{HDLR}
  \by=\beta_{0}\mathbf{1}_{n}+\bX\bbeta+\boldsymbol{\varepsilon},
\end{equation}
where $\bX=(\bx_{1}^{\T},\bx_{2}^{\T},\cdots,\bx_{n}^{\T})^{\T}=(\bx_{(1)},
\bx_{(2)},\cdots,\bx_{(p)})$ is an $n\times p$ design matrix with
$\bx_{i}\in\real^{p}$ for $1\leq{}i\leq{}n$ and $\bx_{(j)}\in\real^{n}$ for
$1\leq{}j\leq{}p$, $\by=(y_{1},y_{2},\cdots,y_{n})^{\T}$ is an $n$-dimensional
response vector, $\bbeta=(\beta_{1},\beta_{2},\cdots, \beta_{p})^{\T}$ is a
$p$-dimensional regression coefficient vector, and
$\boldsymbol{\varepsilon}=(\varepsilon_{1},\varepsilon_{2},\cdots,
\varepsilon_{n})^{\T}$ is an $n$-dimensional error vector. Throughout the paper,
we assume $\varepsilon_{i}$'s are independent and identically distributed and
$\Pr(\varepsilon_{i}\leq0)=\tau$ for some given constant $\tau\in(0,1)$. Under
this model, $\beta_{0}+\bx_{i}^{\T}\bbeta$ is the $\tau$th conditional quantile
of $y_{i}$ given $\bx_{i}$. Suppose the $p$ predictors are divided into $g$
known groups $G_{1},\cdots,G_{g}$ with
$\cup_{\ell=1}^{g}G_{\ell}=\{1,\cdots,p\}$ and $G_{i}\cap{}G_{j}=\emptyset$ for
all $1\leq{}i<j\leq{}g$. Denote by $\bbeta_{G_{\ell}}$ the restriction of the
vector $\bbeta$ to the index set $G_{\ell}$.

To reconstruct the sparsity and group sparsity structure of $\bbeta^{\ast}$, we
consider
\begin{eqnarray}\label{spa-group-QR}
  \min_{\beta_{0},\bbeta}Q_{\tau}(\by-\beta_{0}\mathbf{1}_{n}-\bX\bbeta)
  +\lambda\|\bd\odot\bbeta\|_{1}+\mu\sum_{\ell=1}^{g}
  w_{\ell}\|\bbeta_{G_{\ell}}\|_{2},
\end{eqnarray}
where $Q_{\tau}(\by-\beta_{0}\mathbf{1}_{n}-\bX\bbeta)=\frac{1}{n}\sum_{i=1}^{n}
\rho_{\tau}(y_{i}-\beta_{0}-\bx_{i}^{\T}\bbeta)$ with
$\rho_{\tau}(u)=u(\tau-I\{u\leq0\})$ being the quantile check loss,
$\|\bd\odot\bbeta\|_{1}=\sum_{j=1}^{p}d_{j}|\beta_{j}|$ is the weighted
$L_{1}$-norm of $\bbeta$ with weight vector
$\bd=(d_{1},\cdots,d_{p})^{\T}\geq0$, \(\bw=(w_{1},\ldots,w_{g})^{\T}\) is the
vector of group weights with $w_{\ell}\geq{}0$ for each $\ell=1,\cdots,g$,
$\|\bbeta\|_{2}=\sqrt{\sum_{j=1}^{p}\beta_{j}^{2}}$ is the $L_{2}$-norm of
$\bbeta$, and $\lambda\geq0$ and $\mu\geq0$ are the regularization parameters.
In the sequel, we denote
$h(\bbeta)=\lambda\|\bd\circ\bbeta\|_{1}+\mu\sum_{\ell=1}^{g}w_{\ell}
\|\bbeta_{G_{\ell}}\|_{2}$, where we drop the dependence of \(h\) on
\(\lambda,\mu,\bd,\bw\) and the group partition whenever no confusion arises.
Particular examples of \eqref{spa-group-QR} include penalized quantile
regressions with group lasso and lasso penalties.

Regularization has been widely used in high-dimensional statistical modeling
\citep{bickel2006regularization,efron2007discussion,bickel2009simultaneous,jiang2021robust},
among which the penalized least squares methods are perhaps the most popular, including
the bridge~\citep{frank1993statistical},
lasso~\citep{tibshirani1996regression}, SCAD~\citep{fan2001variable}, adaptive
lasso~\citep{zou2006adaptive}, MCP~\citep{zhang2010nearly}, and related approaches. Despite
the widespread use of penalized least squares in high-dimensional settings, these
methods often fail when faced with outliers or heavy-tailed errors. Accordingly, many recent studies have
focused on sparse penalized quantile regression for robust high-dimensional
data analysis~\citep{belloni2011,wang2012quantile,fan2014adaptive}. Several works have also
investigated computational algorithms for solving penalized quantile regression. For example,
\cite{wu2008coordinate} propose a coordinate descent algorithm for the lasso
penalized least absolute deviation regression that computes a weighted median at
each iteration. \cite{peng2015iterative} design a coordinate descent algorithm
for computing the nonconvex penalized quantile regression.
\cite{yi2017semismooth} consider a semismooth Newton coordinate descent
algorithm to solve the elastic-net penalized quantile regression.
\cite{gu2018admm} construct two variations of the ADMM algorithm for solving the
penalized quantile regression under lasso, adaptive lasso and folded concave
regularization. More recently, \cite{zhang2025wenet} study weighted elastic net
penalized quantile regression and develop an efficient algorithm from the dual
perspective, establishing global convergence and local convergence rates.

In various applications, sparse penalized quantile regression has proven
effective owing to its robustness and sparsity-inducing properties. Nevertheless,
it fails to capture grouping information. For example, in genome-wide
association studies, understanding the biological structure of
a complex disease may require analyzing the joint effects of single nucleotide polymorphisms
within a single gene or across several genes within a biochemical pathway.
In this context, group sparsity rather than individual sparsity is desired.
Group variable selection can be achieved by
incorporating group penalties into regularization-based regression.
\cite{yuan2006model} proposed the group lasso, an extension of the lasso that
performs group-level variable selection in generalized linear
regression models. Group penalized quantile regression has since been widely applied in various fields.
\cite{hashem2016quantile} studied a group lasso penalized quantile regression
for binary responses. \cite{kato2011group} investigated the group lasso
quantile regression and showed theoretical results for the convergence rate
and the oracle property of the estimator. \cite{ouhourane2022group} developed a
group-wise descent algorithm for estimating the parameters for penalized quantile regression with
group lasso, group SCAD and group MCP penalties. \cite{sherwood2022quantile}
proposed a groupwise-majorization-descent (GMD) algorithm to fit the group
lasso penalized Huber approximate quantile regression. Based on the adaptive
idea in \cite{zou2006adaptive}, \cite{wang2008note} proposed the adaptive group
lasso as a way to increase model flexibility and correct estimation bias.
\cite{ciuperca2019adaptive} studied the adaptive group lasso penalized quantile
regression and showed that its estimator possesses asymptotic normality under both a
fixed and a diverging number of groups.
\par Although group lasso penalized quantile regression has been extensively studied in high-dimensional data analysis, it does not ensure sparsity within each group (i.e., it performs group-level but not within-group selection). To achieve sparsity both between and within groups, \cite{friedman2010note} proposed the sparse group lasso, a linear combination of the lasso and group lasso that yields solutions that are sparse both at the group and individual levels. \cite{poignard2020asymptotic} proposed an adaptive sparse group lasso estimator for low-dimensional settings and established its theoretical properties for a class of general convex loss functions. \cite{mendez2021adaptive} extended this to quantile regression in high-dimensional settings. Recent studies also indicate continued interest in quantile-based models with grouped or adaptive sparse regularization in high-dimensional forecasting and risk analysis \citep{chulia2024daily,tao2024green}. Despite these developments, computationally efficient algorithms for fitting sparse group penalized quantile regression have received limited attention in the literature.
\par In this paper, we develop a computationally efficient algorithm for the adaptive sparse group lasso penalized quantile regression. Dual problem formulations have been exploited to improve computational efficiency in a range of high-dimensional estimation problems; for example, \cite{li2016schur} proposed a semi-proximal ADMM based on the Schur complement to solve the dual of a convex quadratic semidefinite program, \cite{zhang2020efficient} designed a semi-smooth Newton augmented Lagrangian method to solve the dual of the sparse group lasso problem, and \cite{li2018highly} proposed a semi-smooth Newton augmented Lagrangian algorithm for the dual of combinatorial convex optimization models, establishing asymptotic superlinear convergence. Motivated by these results and the computational advantages of dual formulations, we propose a novel dual ADMM algorithm (SGL-DADMM) for solving~\eqref{spa-group-QR}. We further establish global convergence for the algorithm.
Numerical experiments confirm that SGL-DADMM efficiently estimates the adaptive sparse group lasso penalized quantile regression and produces solutions of high statistical accuracy. The remainder of this article is organized as follows. Section~\ref{preliminaries} reviews the necessary background. Section~\ref{model and algorithm} presents the model formulation, derives the dual ADMM algorithm, and establishes its global convergence. Section~\ref{section4} describes implementation details, including the computation of $\lambda_{\max}^\alpha$ and the stopping criteria. Simulation studies are presented in Section~\ref{simulation}, and Section~\ref{conclusion} concludes the paper.

\section{Preliminaries}\label{preliminaries}

This section reviews the basic concepts and results underlying the proposed algorithm.
\begin{definition}\label{conjugate}
For any function $f:\real^{d}\to\real$, the convex conjugate of $f$ is
\begin{equation*}
\begin{array}{l}
f^*(\bu)=\sup\limits_{\ba\in \mathbb{R}^d}\left\{{\bu}^T\ba-f(\ba)\right\}, \quad \forall\,\bu\in\mathbb{R}^d.
\end{array}
\end{equation*}
\end{definition}
\begin{definition}\label{proximal}
 For any function $f:\real^{d}\to\real$, the proximal mapping of $f$ is
\[
  \mathrm{prox}_{f}(\ba):=\argmin_{\bu}\Bigl\{f(\bu)
  +\frac{1}{2}\|\bu-\ba\|_{2}^{2}\Bigr\},\qquad\bu\in\real^{d}.
\]
The Moreau identity, used repeatedly below, states that
\begin{equation}
\mathrm{prox}_{\sigma f}(\ba)+\sigma\,\mathrm{prox}_{\sigma^{-1}f^*}(\ba/\sigma)=\ba,\label{Moreau identity}
\end{equation}
for any $\sigma>0$.
\end{definition}

\begin{lemma}\label{quantile-proj}
  For any $\tau\in(0,1),$ consider
  \[
    \min_{\bu\in\real^{d}}\frac{1}{2}\|\bu-\ba\|_{2}^{2},
    \qquad\text{subject to}\qquad-\tau\mathbf{1}
    \leq\bu\leq(1-\tau)\mathbf{1},
  \]
  where $\ba=(a_{1},\cdots,a_{d})^{\T}$ is given. Then, it has a unique solution
  $\hat{\bu}=(\hat{u}_{1},\cdots,\hat{u}_{d})^{\T},$ where
  $\hat{u}_{i}=\max\{-\tau,\min\{(1-\tau),a_{i}\}\}$. For convenience, denote
  \(
    \hat{\bu}=\max\{-\tau\mathbf{1},\min\{(1-\tau)\mathbf{1},\ba\}\}.
  \)
\end{lemma}

Similar to \cite{zhang2020efficient}, we have the following three lemmas.
\begin{lemma}\label{prox-group}
  Denote $\phi_{\mu\circ\bw}(\bbeta)=\mu\sum_{\ell=1}^{g}w_{\ell}
  \|\bbeta_{G_{\ell}}\|_{2}$, where $\mu\geq0$, \(\bw\geq\mathbf{0}\), and
  \(\{G_{1},\ldots,G_{g}\}\) is a non-overlapping group partition of
  \(\{1,\ldots,p\}\). For each $\ell=1,2,\cdots,g$, we define the linear
  operator $\mathcal{P}_{\ell}:\real^{p}\to\real^{|G_{\ell}|}$ by
  $\mathcal{P}_{\ell}\bbeta=\bbeta_{G_{\ell}}$. Then,
  \[
    \text{Prox}_{\phi_{\mu\circ\bw}}(\boldeta)=\sum_{\ell=1}^{g}
    \mathcal{P}_{\ell}^{\T}\boldeta_{\ell}^{\mu\circ\bw},
  \]
  where
  \[
    \begin{aligned}
      \boldeta_{\ell}^{\mu\circ\bw}=\left\lbrace
        \begin{array}{lll}
          \frac{\boldeta_{G_{\ell}}}{\|\boldeta_{G_{\ell}}\|_{2}}
          \max\{\|\boldeta_{G_{\ell}}\|_{2}-\mu{}w_{\ell},0\},~
          &\boldeta_{G_{\ell}}\neq\mathbf{0},\\
          \mathbf{0},& \text{otherwise},
        \end{array}\right.
    \end{aligned}
  \]
  for each $\ell=1,2\cdots,g$.
\end{lemma}
\begin{lemma}\label{prox-ada-lasso}
  Denote $\varphi_{\lambda\circ\bd}(\bbeta)=\lambda\|\bd\odot\bbeta\|_{1}$,
  where $\lambda\geq0$ and \(\bd\geq\mathbf{0}\). Then,
  \[
    \text{Prox}_{\varphi_{\lambda\circ\bd}}(\bxi)=\text{sgn}(\bxi)\odot
    \max\{|\bxi|-\lambda{}\bd,\mathbf{0}\}.
  \]
\end{lemma}
\begin{lemma}\label{ada-lasso-group}
  For any $\lambda,\mu\geq0$ and $\bd,\bw\geq\mathbf{0}$, it follows that
  \[
    \text{Prox}_{\varphi_{\lambda\circ\bd}+\phi_{\mu\circ\bw}}(\bxi)
    =\text{Prox}_{\phi_{\mu\circ\bw}}
    \big(\text{Prox}_{\varphi_{\lambda\circ\bd}}(\bxi)\big).
  \]
\end{lemma}

\section{Algorithm}\label{model and algorithm}
In this section, we introduce the SGL-DADMM algorithm for computing the adaptive sparse group lasso penalized quantile regression estimator and establish its global convergence.
\subsection{Model Analysis}
We introduce an efficient iterative algorithm for solving model~\eqref{spa-group-QR}. Letting $\bz=\by-\beta_0\mathbf{1}-\bX\bbeta$, model~\eqref{spa-group-QR} can be rewritten as
\begin{eqnarray}\label{spa-group-QR1}
  \min_{\beta_{0},\bbeta,\bz}&&Q_{\tau}(\bz)
  +h(\bbeta)\nonumber\\
 \text{s.t.}~&&\bX\bbeta+\beta_0\mathbf{1}-\by-\bz=0.
\end{eqnarray}
The Lagrangian function of \eqref{spa-group-QR1}~is
\begin{equation*}
\begin{array}{l}
\mathscr{L}(\beta_0,\bbeta,\bz,\btheta)=Q_{\tau}(\bz)+h(\bbeta)+\langle\btheta,\bX\bbeta+\bz+\beta_0\mathbf{1}-\by\rangle\text{,}
\end{array}
\end{equation*}
where ${\btheta\in \mathbb{R}^n}$ is the Lagrange multiplier and
\begin{equation*}
\begin{split}
&\max_{\btheta}\inf_{\beta_0,\bbeta,\bz}\mathcal{L}(\beta_0,\bbeta,\bz,\btheta)\\
&=\max_{\btheta}\{-\langle\by,\btheta\rangle+\inf_{\bbeta} \{h(\bbeta)+\langle
\bX^T\btheta,\bbeta\rangle\}+\inf_{\bu}\{Q_{\tau}(\bz)+\langle\btheta,\bz\rangle\}+\inf_{\beta_0}\langle\bm{1}^T\btheta,\beta_0\rangle\}\\
&=\max_{\btheta}\{-\langle\by,\btheta\rangle-\sup_{\bbeta} \{-h(\bbeta)+\langle
-\bX^T\btheta,\bbeta\rangle\}-\sup_{\bz}\{-Q_{\tau}(\bz)+\langle-\btheta,\bz\rangle\}+\inf_{\beta_0}\langle\bm{1}^T\btheta,\beta_0\rangle\}\\
&=\max_{\btheta}\{-\langle\btheta,\by\rangle-h^*(-\bX^T\btheta)-Q_{\tau}^*(-\btheta)\}\\
&~~~~~~~\mathrm{s.t.}~~\bm{1}^T\btheta=0.
\end{split}
\end{equation*}
A straightforward calculation yields the dual problem:
\begin{eqnarray}
  \max_{\btheta,\bu}&&-\langle\by,\btheta\rangle{}-h^{\ast}(\bu)\nonumber\\
  \text{s.t.}&&\bX^{\T}\btheta+\bu=\mathbf{0}\nonumber,\\
                    &&\mathbf{1}^{\T}\btheta=\mathbf{0}\nonumber,\\
                    &&-\tau\mathbf{1}\leq\btheta
                       \leq(1-\tau)\mathbf{1}.\nonumber
\end{eqnarray}
Equivalently, we solve the minimization problem
\begin{eqnarray}\label{dual-prob}
  \min_{\btheta,\bu,\bv}&&\langle\by,\btheta\rangle{}+h^{\ast}(\bu)
                           +\delta_{\mathcal{C}}(\bv)\nonumber\\
  \text{s.t.}&&\bX^{\T}\btheta+\bu=\mathbf{0}\nonumber,\\
                        &&\btheta-\bv=\mathbf{0}\nonumber,\\
                        &&\mathbf{1}^{\T}\btheta=\mathbf{0}.
\end{eqnarray}
where $\mathcal{C}=\{\bv\in\real^{n}:-\tau\mathbf{1}
\leq\bv\leq(1-\tau)\mathbf{1}\}$ and
\[
  \delta_{\mathcal{C}}(\bv)=\left\{
    \begin{array}{lll}
      0, & \text{if}~\bv\in\mathcal{C},\\
      +\infty, & \text{otherwise}.
    \end{array}
  \right.
\]
The Karush-Kuhn-Tucker (KKT) system of the dual problem \eqref{dual-prob} is
\begin{equation}
\left\{ \begin{array}{l}
{\bX\bbeta+\bz+\beta_0\bm{1}-\by=\bm{0}}\text{,}\\
{\bu-Prox_{h^*}(\bbeta+\bu)=0}\text{,}\\
\bv-Prox_{\delta_{\mathcal{C}}}(\bv-\bz)\text{,}\\
{\bX^T\btheta+\bu=\bm{0}}\text{,}\\
{\btheta-\bv=\bm{0}}\text{,}\\
{\bm{1}^T\btheta=0}\text{.}
\label{KKT}
\end{array} \right.
\end{equation}
\subsection{SGL-DADMM Algorithm}
We now present an efficient iterative algorithm for solving the dual problem~\eqref{dual-prob}.

For $\varpi>0$, the augmented Lagrange function related to problem (\ref{dual-prob}) is given by
\[
  \begin{aligned}
    \mathcal{L}_{\varpi}(\btheta,\bu,\bv,\bLambda_{1},\bLambda_{2},\Lambda_{3})
    =&\,\langle\by,\btheta\rangle{}+h^{\ast}(\bu)
    +\delta_{\mathcal{C}}(\bv)-\langle\bLambda_{1},
    \bX^{\T}\btheta+\bu\rangle-\langle\bLambda_{2},
    \btheta-\bv\rangle-\langle\Lambda_{3},
    \mathbf{1}^{\T}\btheta\rangle\\
    &\,+\frac{\varpi}{2}\|\bX^{\T}\btheta+\bu\|_{2}^{2}
    +\frac{\varpi}{2}\|\btheta-\bv\|_{2}^{2}
    +\frac{\varpi}{2}(\mathbf{1}^{\T}\btheta)^{2},
  \end{aligned}
\]
where $\bLambda_{1}\in\mathbb{R}^p$, $\bLambda_{2}\in\mathbb{R}^n$, and $\bLambda_{3}\in\mathbb{R}$ are Lagrange multipliers. Applying ADMM to problem~\eqref{dual-prob} yields
\begin{eqnarray}\label{proximal-ADMM-1}
  \left\lbrace
  \begin{array}{lll}
    \btheta^{k+1}&=&\arg\min\limits_{\btheta}\mathcal{L}_{\varpi}
                     (\btheta,\bu^{k},\bv^{k},
                     \bLambda_{1}^{k},\bLambda_{2}^{k},\Lambda_{3}^k),\\
    \bu^{k+1}&=&\arg\min\limits_{\bu}\mathcal{L}_{\varpi}
                 (\btheta^{k+1},\bu,\bv^k,
                 \bLambda_{1}^{k},\Lambda_{2}^{k},\Lambda_{3}^k),\\
    \bv^{k+1}&=&\arg\min\limits_{\bv}\mathcal{L}_{\varpi}
                 (\btheta^{k+1},\bu^k,\bv,
                 \bLambda_{1}^{k},\Lambda_{2}^{k},\Lambda_{3}^K),\\
    \bLambda_{1}^{k+1}&=&\bLambda_{1}^{k}-\gamma\varpi
                          (\bX^{\T}\btheta^{k+1}+\bu^{k+1}),\\
    \bLambda_{2}^{k+1}&=&\bLambda_{2}^{k}-\gamma\varpi
                          (\btheta^{k+1}-\bv^{k+1}),\\
    \Lambda_{3}^{k+1}&=&\Lambda_{3}^{k}-\gamma\varpi
                         (\mathbf{1}^{\T}\btheta^{k+1}),
  \end{array}\right.
\end{eqnarray}
where $\gamma\in(0,(\sqrt{5}+1)/2)$. The framework in \eqref{proximal-ADMM-1} involves solving four optimization subproblems, of which the solutions are given below in (i)-(iii).
\begin{enumerate}
\item[(i)]$\btheta$-subproblem: 
   \begin{eqnarray}
    \btheta^{k+1}=\arg\min_{\btheta}\langle{}
                     \by-\bX\bLambda_{1}^{k}-\bLambda_{2}^{k}
                     -\Lambda_{3}^{k}\mathbf{1},
                     \btheta\rangle+\frac{\varpi}{2}
                     \|\bX^{\T}\btheta+\bu^{k}\|_{2}^{2}
                     +\frac{\varpi}{2}
                     \|\btheta-\bv^{k}\|_{2}^{2}+\frac{\varpi}{2}
                     (\mathbf{1}^{\T}\btheta)^{2},\nonumber
\end{eqnarray}
Note that $\btheta^{k+1}$ can be obtained by solving the following linear system:
\begin{equation*}
\begin{array}{l}
\sigma(\bX\bX^T+\bI_n+\bm{1}\bm{1}^T)\btheta^{k+1}=\sigma \bv^k+\bX\bLambda_1^k+\bLambda_2^k-\bm{1}\Lambda_3^k-\sigma \bX\bz^k-\by\text{.}
\end{array}
\end{equation*}
\item[(ii)]$\bu$-subproblem: the optimization of $\bu$ can be formulated as
\begin{eqnarray}
  \begin{array}{lll}
\bu^{k+1}&=\arg\min\limits_{\bu} \varpi^{-1}h^{\ast}(\bu)
                 +\frac{1}{2}\|\bu+\bX^{\T}\btheta^{k+1}
                 -\varpi^{-1}\bLambda_{1}^{k}\|_{2}^{2}\nonumber\\
&=\text{Prox}_{h^{\ast}/\varpi}
                 ((\bLambda_{1}^{k}-\varpi\bX^{\T}\btheta^{k+1})/\varpi),\\
\end{array}
\end{eqnarray}
by the Moreau identity~\eqref{Moreau identity}, we have
\[
  \bu^{k+1}=\frac{(\bLambda_{1}^{k}-\varpi\bX^{\T}\btheta^{k+1})
    -\text{Prox}_{\varpi{}h}\big(\bLambda_{1}^{k}
    -\varpi\bX^{\T}\btheta^{k+1}\big)}{\varpi}.
\]
which together with Lemmas \ref{prox-group}--\ref{ada-lasso-group} yields that
\[
\bu^{k+1}=\varpi^{-1}\big(\bLambda_{1}^{k}-\varpi\bX^{\T}
                 \btheta^{k+1}-\text{Prox}_{\phi_{\varpi\mu\circ\bw}}\big(
                 \text{Prox}_{\varphi_{\varpi\lambda\circ\bd}}(\bLambda_{1}^{k}
                 -\varpi\bX^{\T}\btheta^{k+1})\big)\big).
\]
\item[(iii)]$\bv$-subproblem: the optimization of $\bv$ can be simplified to
\begin{eqnarray}
\bv^{k+1}=\arg\min_{\bv}\delta_{\mathcal{C}}(\bv)
                 +\frac{1}{2}\|\bv-\btheta^{k+1}
                 +\varpi^{-1}\bLambda_{2}^{k}\|_{2}^{2},\nonumber
\end{eqnarray}
which together with Lemma \ref{quantile-proj} yields that
\begin{eqnarray}
\bv^{k+1}=\max\{-\tau\mathbf{1},\min\{(1-\tau)\mathbf{1},
                 \btheta^{k+1}-\varpi^{-1}\bLambda_{2}^{k}\}\}.\nonumber
\end{eqnarray}
\end{enumerate}

Combining the above results with the strong duality and saddle point theorems for the convex
optimization problem~\eqref{spa-group-QR}, we arrive at the following ADMM algorithm based on the dual formulation.

\renewcommand{\algorithmicrequire}{\textbf{Input:}} 
\renewcommand{\algorithmicensure}{\textbf{Output:}}
\begin{algorithm}[H]\small
\caption{The SGL-DADMM algorithm for solving the model \eqref{spa-group-QR}.}
\label{slg:admm}
\begin{algorithmic}[111]
\REQUIRE
Data $(\bX,\by)$,
parameters~$\gamma,\lambda,\mu,\tau$.
\ENSURE The iterative sparse solution~$\widehat{\bbeta}$. \\
\STATE{Initialize the algorithm with~$(\btheta^0,\bu^0,\bv^0,\bbeta^0,\bz^0,\beta_0^0)$;}
\STATE{$k \gets 0$;}
\REPEAT
\STATE{Update~$\btheta^{k+1}\gets(\bX\bX^{\T}+\mathbf{1}\mathbf{1}^{\T}+\bI_{n})^{-1}
                      \big(\bv^{k}-\bX\bu^k+\varpi^{-1}
                      \big(\bX\bbeta^{k}+\bLambda_{2}^{k}
                      +\beta_{0}^{k}\mathbf{1}-\by\big)\big)$;}
\STATE{Update~$\bu^{k+1}\gets\varpi^{-1}\big(\bbeta^{k}-\varpi\bX^{\T}\btheta^{k+1}
                  -\text{Prox}_{\phi_{\varpi\mu\circ\bw}}
                  \big(\text{Prox}_{\varphi_{\varpi\lambda\circ\bd}}
                  ({\bbeta}^{k}-\varpi\bX^{\T}\btheta^{k+1})\big)\big)$;}
\STATE{Update~$\bv^{k+1}\gets\max\{-\tau\mathbf{1},\min\{(1-\tau)\mathbf{1},
                  \btheta^{k+1}-\varpi^{-1}\bLambda_{2}^{k}\}\}$;}
\STATE{Update~$     \bbeta^{k+1}\gets\bbeta^{k}-\gamma\varpi(\bX^{\T}\btheta^{k+1}+\bu^{k+1})$;}
\STATE{Update~$\bz^{k+1}\gets\bz^{k}-\gamma\varpi(\btheta^{k+1}-\bv^{k+1})$;}
\STATE{Update~$\beta_{0}^{k+1}\gets\beta_{0}^{k}-\gamma\varpi(\mathbf{1}^{\T}\btheta^{k+1})$;}
\STATE{Update $k\gets k+1$};
\UNTIL{the convergence criterion is met}
\RETURN $\widehat{\bbeta}=\bbeta^k$.
\end{algorithmic}
\end{algorithm}

 Basically, we identify
\(\bLambda_{1}\),\(\bLambda_{2}\), and \(\Lambda_{3}\) with \(\bbeta\), \(\bz\), and \(\beta_{0}\),
respectively. Note that when \(\gamma=1\), the update for
\(\bbeta\) simplifies to
\begin{equation*}
  \begin{split}
    \bbeta^{k+1}&=\bbeta^{k}-\varpi\bX^{\T}\btheta^{k+1}-\varpi\bu^{k}\\
    &=\text{Prox}_{\phi_{\varpi\mu\circ\bw}}\big(
    \text{Prox}_{\varphi_{\varpi\lambda\circ\bd}}
    (\bbeta^{k}-\varpi\bX^{\T}\btheta^{k+1})\big),
  \end{split}
\end{equation*}
which now possesses the variable/group sparsity structure.

\subsection{Convergence Analysis}
We now establish convergence properties for Algorithm~\ref{slg:admm}. Let
$\bxi=\begin{bmatrix}{\bu}\\{\bv}\\0\end{bmatrix}\in\mathbb{R}^{p+n+1}$, ${{\bLambda}}=\begin{bmatrix}{\bbeta}\\{\bz}\\{\beta_0}\end{bmatrix}\in\mathbb{R}^{p+n+1}$
and ${\bA}={\begin{bmatrix}{\bX}_{p\times n}^T \\ {\bI}_{n}\\\bm{1}_n^T\end{bmatrix}}$, $
{\bm{B}}={\begin{bmatrix}{\bI}_p&{0}& 0\\{0}& -{\bI}_n& 0\\ {0}&{0}& \bm{1}\end{bmatrix}}$.
Define $g(\bxi)=h^*(\bu)+\delta_{\mathcal{C}}(\bv)$. Then, the dual problem \eqref{dual-prob} can be reformulated into the following two standard convex optimization subproblems:
\begin{equation}
\begin{array}{l}
\mathop {\min}\limits_{{\btheta},{\bxi}}~\langle \by,\btheta\rangle+g(\bxi)\\
\mathrm{s.t.}~~~ \bA{\btheta}+\bm{B}{\bxi}=0\text{.}\label{shoulianduiou2}
\end{array}
\end{equation}

Following the proof strategy of \cite{2015An} (Theorem~5.1), we establish the global convergence of the SGL-DADMM algorithm as follows; the detailed proof is provided in the appendix.
\begin{theorem}\label{theorem2}
Let ${\bgamma\in\left(0,(1+\sqrt{5})/2\right)}$, and $\left\{({\btheta}^k,{\bu}^k,{\bv}^k,{\bbeta}^k,{\bz}^k,\beta_0^k)\right\}$ be the sequence generated by the SGL-DADMM algorithm. Then, $\left\{\left({\btheta}^k,{\bu}^k,{\bv}^k\right)\right\}$ converges to the optimal solution of the dual problem \eqref{shoulianduiou2}, and $\left\{\left({\bbeta}^k,{\bz}^k,\beta_0^k\right)\right\}$ converges to the optimal solution of its primal problem.
\end{theorem}

\section{Implementation}\label{section4}
This section describes implementation details for the proposed algorithm, including the computation of $\lambda_{\mathrm{max}}^\alpha$ (the smallest
\(\lambda\) for which all coefficients in \(\bbeta\) are zero) and the stopping criteria.
\subsubsection*{Computation of $\lambda_{max}^\alpha$}

\par Similar to \texttt{glmnet}, we work with the following parameterization of the adaptive sparse
group lasso penalized quantile regression
\begin{equation*}
  \min_{\beta_{0},\bbeta}\frac{1}{n}\sum_{i=1}^{n}
  \rho_{\tau}(y_{i}-\beta_{0}-\bx_{i}^{\T}\bbeta)
  +\lambda\Bigl[(1-\alpha)\sum_{j=1}^{p}d_{j}|\beta_{j}|
  +\alpha\sum_{\ell=1}^{g}w_{\ell}\|\bbeta_{G_{\ell}}\|_{2}\Bigr]
\end{equation*}
for some \(\alpha\in[0,1]\). 
\begin{itemize}
\item When \(\alpha=0\), the problem reduces to the (weighted) lasso penalized
  quantile regression and one choice of such \(\lambda\) is
  \begin{equation*}
    \lambda_{\max}^{0}=\max_{\substack{1\leq{}j\leq{}p\\d_{j}\neq{}0}}
    \frac{1}{d_{j}}\biggl\{\biggl|\frac{2\tau-1}{2n}
    \sum_{i=1}^{n}x_{ij}+\frac{1}{2n}
    \sum_{i\not\in\mathcal{Z}}\text{sgn}(r_{i})x_{ij}\biggr|+
    \frac{1}{2n}\sum_{i\in\mathcal{Z}}|x_{ij}|\biggr\},
  \end{equation*}
  where \(\mathcal{Z}=\{1\leq{}i\leq{}n:r_{i}=0\}\) and
  \(r_{i}=y_{i}-\tilde{\beta}_{0}\) with \(\tilde{\beta}_{0}\) being the
  \(\tau\)th sample quantile of the \(y_{i}\)'s.
\item When \(\alpha=1\), the problem reduces to the (weighted) group lasso
  penalized quantile regression and one choice of such \(\lambda\) is
  \begin{equation*}
    \lambda_{\max}^{1}=\max_{\substack{1\leq{}\ell\leq{}L\\w_{\ell}\neq{}0}}
    \frac{1}{w_{\ell}}\biggl\{\biggl\|\frac{2\tau-1}{2n}
    \sum_{i=1}^{n}\bx_{i,G_{\ell}}+\frac{1}{2n}
    \sum_{i\not\in\mathcal{Z}}\text{sgn}(r_{i})\bx_{i,G_{\ell}}
    \biggr\|_{2}+\frac{1}{2n}\sum_{i\in\mathcal{Z}}
    \|\bx_{i,G_{\ell}}\|_{2}\biggr\}.
  \end{equation*}
\item When \(0<\alpha<1\), this is the general sparse group lasso penalized
  quantile regression and one choice of such \(\lambda\) is given by
  \begin{equation*}
    \begin{split}
      \lambda_{\max}^{\alpha}=\min\Biggl\{
      &\max_{\substack{1\leq{}j\leq{}p\\d_{j}\neq{}0}}
      \frac{1}{(1-\alpha)d_{j}}\biggl(\biggl|\frac{2\tau-1}{2n}
      \sum_{i=1}^{n}x_{ij}+\frac{1}{2n}
      \sum_{i\not\in\mathcal{Z}}\text{sgn}(r_{i})x_{ij}\biggr|+
      \frac{1}{2n}\sum_{i\in\mathcal{Z}}|x_{ij}|\biggr),\\
      &\max_{\substack{1\leq{}\ell\leq{}L\\w_{\ell}\neq{}0}}
      \frac{1}{\alpha{}w_{\ell}}\biggl(\biggl\|\frac{2\tau-1}{2n}
      \sum_{i=1}^{n}\bx_{i,G_{\ell}}+\frac{1}{2n}
      \sum_{i\not\in\mathcal{Z}}\text{sgn}(r_{i})\bx_{i,G_{\ell}}
      \biggr\|_{2}+\frac{1}{2n}\sum_{i\in\mathcal{Z}}
      \|\bx_{i,G_{\ell}}\|_{2}\biggr)\Biggr\}.
    \end{split}
  \end{equation*}
\end{itemize}
The value of \(\lambda_{\max}^{\alpha}\) may be slightly larger than
required depending on the size of \(\mathcal{Z}\), but it guarantees that all coefficients in \(\bbeta\) are zero.

We need to compute the inverse matrix of
\((\bI_{n}+\bX\bX^{\T}+\mathbf{1}_{n}\mathbf{1}_{n}^{\T})\). This can be
computed directly when \(n\) is moderate. When \(n\) is large, we turn to \(p\).
Here we assume \(\bX\) is column-centered (i.e., every
column \(\bx_{(j)}\) has zero mean, which is always achievable since an intercept is included).
If \(p\) is small, we can apply the Woodbury
identity
\begin{equation*}
  (\bI_{n}+\bX\bX^{\T}+\mathbf{1}_{n}\mathbf{1}_{n}^{\T})^{-1}
  =\bI_{n}-\bX(\bI_{p}+\bX^{\T}\bX)^{-1}\bX^{\T}-\frac{1}{n+1}
  \mathbf{1}_{n}\mathbf{1}_{n}^{\T},
\end{equation*}
where the major computation is for the inverse of matrix \((\bI_{p}+\bX^{\T}\bX)\).
When \(p\) is also large, directly computing either the \(n \times n\) or the \(p \times p\) inverse matrix becomes prohibitively expensive. In such scenarios, it is more efficient to solve the associated linear system iteratively. Since the matrix \(\bI_{n}+\bX\bX^{\T}+\mathbf{1}_{n}\mathbf{1}_{n}^{\T}\) is symmetric and positive definite, the Conjugate Gradient (CG) method is a natural and well-suited choice. Furthermore, its convergence can be significantly accelerated by employing a suitable preconditioner, such as a diagonal preconditioner based on the matrix's main diagonal or an approximation derived from its structure. This approach avoids the computational and memory burden of explicit matrix inversion, making it practical for large-scale problems.


\subsubsection*{Stopping Criteria}

Denote
\[
  \epsilon^{\text{pri}}=\epsilon_{1}\sqrt{p+n+1}+\epsilon_{2}
  \max\biggl\{\sqrt{\|\bX^{\T}\btheta^{k+1}\|_{2}^{2}+
    \|\btheta^{k+1}\|_{2}^{2}+(\mathbf{1}^{\T}\btheta^{k+1})^{2}},
  \sqrt{\|\bu^{k+1}\|_{2}^{2}+\|\bv^{k+1}\|_{2}^{2}}\biggr\}
\]
and
\[
  \epsilon^{\text{dual}}=\sqrt{n}\epsilon_{1}+\epsilon_{2}
  \|\bX\beta^{k+1}+\bLambda_{2}^{k+1}+\beta_{0}^{k+1}\mathbf{1}\|.
\]
The Algorithm \ref{slg:admm} is terminated either when the sequence meets the following
criterion

\[
  \sqrt{\|\bX^{\T}\btheta^{k+1}+\bu^{k+1}\|_{2}^{2}
    +\|\btheta^{k+1}-\bv^{k+1}\|_{2}^{2}+(\mathbf{1}^{\T}\btheta^{k+1})^{2}}
  \leq\epsilon^{\text{pri}}
\]
and
\[
  \varpi\|\bX(\bu^{k+1}-\bu^{k})-(\bv^{k+1}-\bv^k)\|_{2}\leq\epsilon^{\text{dual}},
\]
where typical choices are $\epsilon_{1}=10^{-3}$ and $\epsilon_{2}=10^{-3},$ or
when the number of ADMM iterations exceeds a certain number.
\section{Simulations}\label{simulation}
We compare the proposed SGL-DADMM estimator with the following alternative methods:
\begin{enumerate}
\item sparsegl: Using the R package \texttt{sparsegl} (\cite{liang2022sparsegl}), the GMD algorithm is employed to solve sparse group lasso penalized least squares problem.
\item hrqglas: Utilizing the R package \texttt{hrqglas} (\cite{sherwood2022quantile}), the HAQ-GMD algorithm is applied to solve the group lasso penalized quantile regression problem.
\item GPQR: With the R package \texttt{GPER} (\cite{ouhourane2022group}), the GPQR algorithm is utilized to solve the group lasso penalized quantile regression problem.
\item hqreg: Employing the R package \texttt{hqreg} (\cite{yi2017semismooth}), the semi-smooth Newton algorithm is used to solve the Huber approximate quantile regression problem with Lasso penalty.
\item SQR: Using the R package \texttt{SQR} (\cite{SQR}), the coordinate descent algorithm is applied to solve the pseudo quantile regression problem with Lasso penalty.
\end{enumerate}
\subsection{Timing Comparisons}\label{Timing Comparisons}

Following \cite{SQR}, we conduct simulations with sample size $n = 100$ and dimensions $p \in \{500, 1000\}$. The true parameters are set to \( \bbeta^* = (3, 3, 3, 3, 2, 2, 2, 2, \\-1, -1, -1, -1, \dots, 0_{p-12})^T \). The random variables \( \bZ_1, \dots, \bZ_p \) are simulated following the standard normal distribution. Covariates \( \bx_{(1)}, \dots, \bx_{(p)} \) are generated according to Equation \eqref{17}:
\begin{equation}\label{17}
    \bx_{(j)} =
    \begin{cases}
        \bZ_1 + \varepsilon_j, & j = 1, \dots, 4, \\
        \bZ_2 + \varepsilon_j, & j = 5, \dots, 8, \\
        \bZ_3 + \varepsilon_j, & j = 9, \dots, 12, \\
        \bZ_{j-12}, & j = 13, \dots, p,
    \end{cases}
\end{equation}
where \( \varepsilon_j \overset{i.i.d}{\sim} N(0,0.01), j = 1, \dots, 12 \). The response variable \( \by \) is generated by the following linear regression model:
\[
\by = \sum_{j=1}^{p} \beta_j^* \bx_{(j)} + \boldsymbol{\varepsilon},
\]
where the error term \( \boldsymbol{\varepsilon} \) considers three distributions: (1) normal distribution with mean 0 and standard deviation 3, denoted as \( N(0,3^2) \); (2) Laplace distribution, denoted as \( L(0,1) \); (3) \( t \) distribution with 4 degrees of freedom, denoted as \( t_4 \).

The performance of each method is evaluated using the following measures:
\begin{enumerate}
  \item MSE: the mean squared error \(\left(1/p\|\hat{\bbeta}-\bbeta^*\|^2_2\right)\).
  \item MAE:~the mean absolute error \( \left(1/p\| \hat{\bbeta} - \bbeta^* \|_1\right) \).
  \item time:~the running time of the algorithm (seconds).
\end{enumerate}

To ensure a fair comparison, this section focuses on the computational time and estimation accuracy of SGL-DADMM relative to HAQ-GMD and GPQR for the group lasso penalized quantile regression model. Results from 100 independent replications are summarized in Tables~\ref{AA1}--\ref{AA3}, with standard errors shown in subscript.

Tables~\ref{AA1}--\ref{AA3} report the MSE, MAE, and computational time of SGL-DADMM, HAQ-GMD, and GPQR under different error distributions. Across all settings, the SGL-DADMM algorithm is substantially faster than both GPQR and HAQ-GMD. For instance, when errors follow a normal distribution, the HAQ-GMD runtime ranges from approximately $1.66$ to $6.54$ seconds and the GPQR runtime from $0.14$ to $0.21$ seconds, whereas SGL-DADMM requires at most $0.02$ seconds. Analogous gains are observed under Laplace and $t$ errors.

\begin{table}[htbp]
  \begin{center}
\footnotesize
    \caption{Simulation Results of the Algorithm When $\boldsymbol{\varepsilon}\sim N(0,3^2)$.}\label{AA1}
    \setlength{\tabcolsep}{4pt}  
    \begin{tabular}{cccccccc} 
\toprule
&&SGL-DADMM&HAQ-GMD&GPQR&SGL-DADMM&HAQ-GMD&GPQR\\
\cmidrule(lr){3-5}\cmidrule(lr){6-8}
&~&~&$p=500$&&~&$p=1000$&\\
\midrule
$\tau=0.25$&MSE&$0.0471_{(0.0160)}$&$0.1275_{(0.0625)}$&$0.1284_{(0.0678)}$&$0.0211_{(0.0078)}$&$0.0581_{(0.0289)}$&$0.0552_{(0.0346)}$\\
&MAE&$0.0573_{(0.0065)}$&$0.0529_{(0.0080)}$&$0.0523_{(0.0083)}$&$0.0248_{(0.0030)}$&$0.0258_{(0.0040)}$&$0.0247_{(0.0046)}$\\
&time&$0.01_{(0.01)}$&$1.93_{(0.89)}$&$0.15_{(0.08)}$&$0.02_{(0.01)}$&$5.68_{(2.18)}$&$0.20_{(0.08)}$\\
\midrule
$\tau=0.5$&MSE&$0.0471_{(0.0160)}$&$0.1292_{(0.0618)}$&$0.1255_{(0.0703)}$&$0.0211_{(0.0078)}$&$0.0576_{(0.0315)}$&$0.0537_{(0.0352)}$\\
&MAE&$0.0573_{(0.0065)}$&$0.0533_{(0.0079)}$&$0.0513_{(0.0090)}$&$0.0248_{(0.0030)}$&$0.0258_{(0.0040)}$&$0.0242_{(0.0047)}$\\
&time&$0.01_{(0.01)}$&$1.66_{(0.63)}$&$0.14_{(0.08)}$&$0.02_{(0.01)}$&$4.13_{(2.23)}$&$0.19_{(0.07)}$\\
\midrule
$\tau=0.75$&MSE&$0.0471_{(0.0160)}$&$0.1315_{(0.0637)}$&$0.1270_{(0.0693)}$&$0.0211_{(0.0078)}$&$0.0568_{(0.0312)}$&$0.0552_{(0.0343)}$\\
&MAE&$0.0678_{(0.0065)}$&$0.0534_{(0.0085)}$&$0.0519_{(0.0093)}$&$0.0248_{(0.0030)}$&$0.0257_{(0.0043)}$&$0.0249_{(0.0045)}$\\
&time&$0.01_{(0.01)}$&$1.72_{(0.53)}$&$0.14_{(0.08)}$&$0.02_{(0.01)}$&$6.54_{(2.57)}$&$0.21_{(0.09)}$\\
\bottomrule
    \end{tabular}
  \end{center}
\end{table}

With respect to MSE and MAE, SGL-DADMM consistently achieves the lowest MSE across quantile levels, while the MAE values of the three methods are broadly comparable. In summary, SGL-DADMM requires substantially less computational time than HAQ-GMD and GPQR while yielding more accurate estimates.

\begin{table}[htbp]
  \begin{center}
\footnotesize
    \caption{Simulation Results of the Algorithm When $\boldsymbol{\varepsilon}\sim L(0,1)$.}\label{AA2}
    \setlength{\tabcolsep}{4pt}  
    \begin{tabular}{cccccccc} 
\toprule
&&SGL-DADMM&HAQ-GMD&GPQR&SGL-DADMM&HAQ-GMD&GPQR\\
\cmidrule(lr){3-5}\cmidrule(lr){6-8}
&~&~&$p=500$&&~&$p=1000$&\\
\midrule
$\tau=0.25$&MSE&$0.0281_{(0.0108)}$&$0.1122_{(0.0835)}$&$0.1101_{(0.0837)}$&$0.0125_{(0.0046)}$&$0.0541_{(0.0393)}$&$0.0532_{(0.0404)}$\\
&MAE&$0.0308_{(0.0050)}$&$0.0467_{(0.0120)}$&$0.0457_{(0.0128)}$&$0.0137_{(0.0024)}$&$0.0233_{(0.0057)}$&$0.0226_{(0.0060)}$\\
&time&$0.01_{(0.01)}$&$2.53_{(1.94)}$&$0.24_{(0.14)}$&$0.03_{(0.01)}$&$7.38_{(4.74)}$&$0.29_{(0.14)}$\\
\midrule
$\tau=0.5$&MSE&$0.0281_{(0.0108)}$&$0.1092_{(0.0861)}$&$0.1071_{(0.0857)}$&$0.0125_{(0.0046)}$&$0.0507_{(0.0416)}$&$0.0516_{(0.0413)}$\\
&MAE&$0.0308_{(0.0050)}$&$0.0454_{(0.0129)}$&$0.0446_{(0.0133)}$&$0.0137_{(0.0024)}$&$0.0220_{(0.0063)}$&$0.0221_{(0.0061)}$\\
&time&$0.01_{(0.01)}$&$1.60_{(0.78)}$&$0.16_{(0.10)}$&$0.03_{(0.01)}$&$4.42_{(2.53)}$&$0.18_{(0.08)}$\\
\midrule
$\tau=0.75$&MSE&$0.0281_{(0.0108)}$&$0.1119_{(0.0820)}$&$0.1114_{(0.0823)}$&$0.0125_{(0.0046)}$&$0.0539_{(0.0400)}$&$0.0533_{(0.0402)}$\\
&MAE&$0.0308_{(0.0050)}$&$0.0469_{(0.0116)}$&$0.0458_{(0.0124)}$&$0.0137_{(0.0024)}$&$0.0231_{(0.0058)}$&$0.0228_{(0.0058)}$\\
&time&$0.01_{(0.01)}$&$1.83_{(0.68)}$&$0.16_{(0.09)}$&$0.03_{(0.02)}$&$6.04_{(2.51)}$&$0.19_{(0.09)}$\\
\bottomrule
    \end{tabular}
  \end{center}
\end{table}

\begin{table}[htbp]
  \begin{center}
\footnotesize
    \caption{Simulation Results of the Algorithm When $\boldsymbol{\varepsilon}\sim t_4$.}\label{AA3}
    \setlength{\tabcolsep}{4pt}  
    \begin{tabular}{cccccccc} 
\toprule
&&SGL-DADMM&HAQ-GMD&GPQR&SGL-DADMM&HAQ-GMD&GPQR\\
\cmidrule(lr){3-5}\cmidrule(lr){6-8}
&~&~&$p=500$&&~&$p=1000$&\\
\midrule
$\tau=0.25$&MSE&$0.0280_{(0.0116)}$&$0.1168_{(0.0883)}$&$0.1161_{(0.0904)}$&$0.0128_{(0.0054)}$&$0.0465_{(0.0384)}$&$0.0457 _{(0.0391)}$\\
&MAE&$0.0310_{(0.0059)}$&$0.0478_{(0.0124)}$&$0.0466_{(0.0127)}$&$0.0136_{(0.0028)}$&$0.0218_{(0.0055)}$&$0.0210_{(0.0059)}$\\
&time&$0.01_{(0.01)}$&$2.01_{(0.079)}$&$0.24_{(0.14)}$&$0.03_{(0.01)}$&$7.06_{(2.61)}$&$0.33_{(0.14)}$\\
\midrule
$\tau=0.5$&MSE&$0.0280_{(0.0116)}$&$0.1128_{(0.0940)}$&$0.1138_{(0.0913)}$&$0.0128_{(0.0054)}$&$0.0439_{(0.0398)}$&$0.0439_{(0.0397)}$\\
&MAE&$0.0382_{(0.0059)}$&$0.0458_{(0.0135)}$&$0.0457_{(0.0132)}$&$0.0136_{(0.0028)}$&$0.0208_{(0.0060)}$&$0.0204_{(0.0061)}$\\
&time&$0.01_{(0.01)}$&$1.57_{(0.62)}$&$0.18_{(0.13)}$&$0.02_{(0.01)}$&$4.46_{(2.67)}$&$0.21_{(0.10)}$\\
\midrule
$\tau=0.75$&MSE&$0.0280_{(0.0116)}$&$0.1156_{(0.0898)}$&$0.1173_{(0.0881)}$&$0.0128_{(0.0054)}$&$0.0485_{(0.0371)}$&$0.0465_{(0.0386)}$\\
&MAE&$0.0310_{(0.0059)}$&$0.0473_{(0.0127)}$&$0.0471_{(0.0125)}$&$0.0136_{(0.0028)}$&$0.0224_{(0.0053)}$&$0.0213_{(0.0058)}$\\
&time&$0.01_{(0.01)}$&$2.05_{(0.72)}$&$0.16_{(0.09)}$&$0.03_{(0.01)}$&$6.82_{(2.80)}$&$0.22_{(0.10)}$\\
\bottomrule
    \end{tabular}
  \end{center}
\end{table}

\subsection{Finite-Sample Performance}\label{Finite-Sample Performance}
This section examines the finite-sample performance of the proposed method. The simulation design follows \cite{peng2015iterative}. Specifically, we draw $(\tilde{\bX}_1,\tilde{\bX}_2,\dots,\tilde{\bX}_q)$ from a multivariate normal distribution $N_q(\bm{0}_q,\Sigma)$ with $\Sigma=(\sigma_{jk})_{q\times q}$ and $\sigma_{jk}=0.5^{|j-k|}$. We set $\bx_{(j)}=\tilde{\bX}_j$ for $j=2,\dots,q$ and $\bx_{(1)}=\Phi(\tilde{\bX}_1)$. Each variable indexed in $\{6,12,15,20\}$ is represented through a third-order polynomial. The response variable is generated from the following regression model:
\begin{equation*}
  \by=\bx_{(6)}+\bx_{(6)}^2+\bx_{(6)}^3+\frac{1}{3}\bx_{(12)}-\bx_{(12)}^2+\frac{2}{3}\bx_{(12)}^3+\frac{1}{2}\bx_{(15)}-\bx_{(15)}^2+\frac{1}{2}\bx_{(15)}^3+\bx_{(20)}+\bx_{(20)}^2+\bx_{(20)}^3+\bx_{(1)}\epsilon.
\end{equation*}
The errors $\epsilon$ follow one of three distributions: (i) homoscedastic normal, $\epsilon\sim N(0,2^2)$; (ii) heteroscedastic normal, $\epsilon=\bx_{(1)}\epsilon^*$ with $\epsilon^*\sim N(0,3^2)$; (iii) asymmetric heteroscedastic, $\epsilon=\bx_{(1)}\epsilon^{**}$ with $\epsilon^{**}\sim \chi^2(3)$. Data are generated with $n=300$ and $q\in\{100,300,500\}$. In all penalized models, $\{\bx_{(j)},\bx_{(j)}^2,\bx_{(j)}^3\}$ is treated as a single group, yielding a final design matrix with $p=3q$ predictors. Results are averaged over 100 replications.

In addition to MSE and MAE (Section~\ref{Timing Comparisons}), estimator performance is assessed by GFP (the number of truly zero coefficients estimated as non-zero) and GFN (the number of truly non-zero coefficients estimated as zero). Mean values over 100 replications for all six methods are reported in Tables~\ref{BB1}--\ref{BB3}, where the proposed method is denoted AGSLQR.

\begin{table}[htbp]
\begin{center}
\small
\captionsetup{skip=5pt}
\setlength{\tabcolsep}{3pt}  
\caption{Simulation Results of Different Methods When Errors $\epsilon$ Follow Normal Distribution.}\label{BB1}
\begin{tabular}{cccccccc} 
\toprule
    &    &ASGLQR&spqrsegl&hrqglas&GPQR&hqreg&SQR\\
\midrule
&MSE&$0.0045_{(0.0024)}$&$0.0070_{(0.0009)}$&$0.0063_{(0.0019)}$&$0.0068_{(0.0008)}$&$0.0080_{(0.0035)}$&$0.0060_{(0.0017)}$\\
$n=300$&MAE&$0.0112_{(0.0026)}$&$0.0138_{(0.0010)}$&$0.0134_{(0.0018)}$&$0.0135_{(0.0009)}$&$0.0251_{(0.0054)}$&$0.0182_{(0.0021)}$\\
$p=300$&GFP&$0.0120_{(0.0103)}$&$0.0126_{(0.0103)}$&$0.0740_{(0.0560)}$&$0.0128_{(0.0104)}$&$0.1670_{(0.0452)}$&$0.2314_{(0.0347)}$\\
&GFN&$0.0046_{(0.0036)}$&$0_{(0)}$&$0_{(0)}$&$0_{(0)}$&$0.0003_{(0.0010)}$&$0.0060_{(0.0024)}$\\
\midrule
&MSE&$0.0021_{(0.0007)}$&$0.0024_{(0.0003)}$&$0.0023_{(0.0003)}$&$0.0023_{(0.0003)}$&$0.0055_{(0.0022)}$&$0.0029_{(0.0006)}$\\
$n=300$&MAE&$0.0046_{(0.0009)}$&$0.0047_{(0.0003)}$&$0.0052_{(0.0007)}$&$0.0046_{(0.0003)}$&$0.0151_{(0.0034)}$&$0.0093_{(0.0010)}$\\
$p=900$&GFP&$0.0056_{(0.0045)}$&$0.0087_{(0.0054)}$&$0.0416_{(0.0337)}$&$0.0077_{(0.0053)}$&$0.1200_{(0.0360)}$&$0.1760_{(0.0215)}$\\
&GFN&$0.0022_{(0.0011)}$&$0_{(0)}$&$0_{(0)}$&$0_{(0)}$&$0.0005_{(0.0008)}$&$0.0031_{(0.0008)}$\\
\midrule
&MSE&$0.0014_{(0.0004)}$&$0.0014_{(0.0002)}$&$0.0016_{(0.0004)}$&$0.0014_{(0.0003)}$&$0.0040_{(0.0016)}$&$0.0021_{(0.0002)}$\\
$n=300$&MAE&$0.0029_{(0.0005)}$&$0.0028_{(0.0002)}$&$0.0032_{(0.0004)}$&$0.0028_{(0.0003)}$&$0.0113_{(0.0024)}$&$0.0064_{(0.0005)}$\\
$p=1500$&GFP&$0.0030_{(0.0021)}$&$0.0076_{(0.0036)}$&$0.0358_{(0.0280)}$&$0.0069_{(0.0037)}$&$0.0988_{(0.0319)}$&$0.1399_{(0.0139)}$\\
&GFN&$0.0015_{(0.0006)}$&$0_{(0)}$&$0_{(0.0002)}$&$0_{(0)}$&$0.0005_{(0.0009)}$&$0.0023_{(0.0004)}$\\
\bottomrule
    \end{tabular}
  \end{center}
\end{table}

\begin{table}[htbp]
\begin{center}
\small
\captionsetup{skip=5pt}
\setlength{\tabcolsep}{3pt}  
\caption{Simulation Results of Different Methods When Errors $\epsilon$ Follow Heteroscedastic Distribution.}\label{BB2}
\begin{tabular}{cccccccc} 
\toprule
    &    &ASGLQR&spqrsegl&hrqglas&GPQR&hqreg&SQR\\
\midrule
&MSE&$0.0026_{(0.0019)}$&$0.0069_{(0.0008)}$&$0.0041_{(0.0015)}$&$0.0067_{(0.0008)}$&$0.0048_{(0.0023)}$&$0.0038_{(0.0013)}$\\
$n=300$&MAE&$0.0075_{(0.0026)}$&$0.0137_{(0.0009)}$&$0.0107_{(0.0019)}$&$0.0134_{(0.0009)}$&$0.0171_{(0.0041)}$&$0.0133_{(0.0019)}$\\
$p=300$&GFP&$0.0039_{(0.0053)}$&$0.0100_{(0.0103)}$&$0.0897_{(0.0572)}$&$0.0092_{(0.0096)}$&$0.1389_{(0.0414)}$&$0.2062_{(0.0309)}$\\
&GFN&$0.0027_{(0.0028)}$&$0_{(0)}$&$0_{(0)}$&$0_{(0)}$&$0_{(0.0003)}$&$0.0046_{(0.0022)}$\\
\midrule
&MSE&$0.0016_{(0.0007)}$&$0.0023_{(0.0003)}$&$0.0019_{(0.0006)}$&$0.0023_{(0.0003)}$&$0.0033_{(0.0016)}$&$0.0023_{(0.0005)}$\\
$n=300$&MAE&$0.0036_{(0.0008)}$&$0.0046_{(0.0003)}$&$0.0043_{(0.0007)}$&$0.0045_{(0.0003)}$&$0.0111_{(0.0031)}$&$0.0075_{(0.0010)}$\\
$p=900$&GFP&$0.0025_{(0.0026)}$&$0.0065_{(0.0050)}$&$0.0483_{(0.0335)}$&$0.0061_{(0.0051)}$&$0.1142_{(0.0339)}$&$0.1640_{(0.0218)}$\\
&GFN&$0.0018_{(0.0010)}$&$0_{(0)}$&$0_{(0)}$&$0_{(0)}$&$0.0002_{(0.0004)}$&$0.0025_{(0.0008)}$\\
\midrule
&MSE&$0.0011_{(0.0004)}$&$0.0014_{(0.0002)}$&$0.0013_{(0.0004)}$&$0.0014_{(0.0002)}$&$0.0029_{(0.0014)}$&$0.0018_{(0.0003)}$\\
$n=300$&MAE&$0.0024_{(0.0005)}$&$0.0028_{(0.0002)}$&$0.0028_{(0.0004)}$&$0.0027_{(0.0002)}$&$0.0089_{(0.0023)}$&$0.0054_{(0.0005)}$\\
$p=1500$&GFP&$0.0013_{(0.0015)}$&$0.0052_{(0.0035)}$&$0.0354_{(0.0255)}$&$0.0048_{(0.0034)}$&$0.0926_{(0.0297)}$&$0.1298_{(0.0122)}$\\
&GFN&$0.0013_{(0.0006)}$&$0_{(0)}$&$0_{(0)}$&$0_{(0)}$&$0.0003_{(0.0007)}$&$0.0021_{(0.0005)}$\\
\bottomrule
    \end{tabular}
  \end{center}
\end{table}

Tables~\ref{BB1}--\ref{BB3} report four metrics for six methods (ASGLQR, sparsegl, hrqglas, GPQR, hqreg, and SQR) across three error distributions and varying $(n,p)$. The key findings are as follows.

\textit{Prediction accuracy.} ASGLQR achieves the lowest or near-lowest MSE and MAE in nearly all settings, demonstrating superior predictive performance. hrqglas also performs well, while hqreg exhibits substantially higher MSE and MAE. sparsegl and GPQR are broadly comparable.

\textit{Variable selection.} The false positive rates (GFP) of ASGLQR, sparsegl, and GPQR remain at very low levels across all settings, indicating strong capacity to identify the sparsity pattern. In contrast, hrqglas, hqreg, and SQR show considerably higher GFP, with SQR having the highest false positive rate throughout. With respect to false negatives (GFN), sparsegl, hrqglas, and GPQR achieve zero GFN, and hqreg maintains near-zero GFN; ASGLQR and SQR have slightly higher but still acceptable GFN values.

As sample size and dimensionality increase, all methods become more stable, as evidenced by decreasing standard deviations. Overall, ASGLQR achieves the best prediction accuracy while maintaining competitive variable selection performance.

\begin{table}[htbp]
\begin{center}
\small
\captionsetup{skip=5pt}
\setlength{\tabcolsep}{3pt}  
\caption{Simulation Results of Different Methods When Errors $\epsilon$ Follow Asymmetric Heteroscedastic Distribution.}\label{BB3}
\begin{tabular}{cccccccc} 
\toprule
    &    &ASGLQR&spqrsegl&hrqglas&GPQR&hqreg&SQR\\
\midrule
&MSE&$0.0032_{(0.0019)}$&$0.0071_{(0.0009)}$&$0.0030_{(0.0014)}$&$0.0067_{(0.0009)}$&$0.0178_{(0.0042)}$&$0.0032_{(0.0011)}$\\
$n=300$&MAE&$0.0093_{(0.0023)}$&$0.0134_{(0.0011)}$&$0.0096_{(0.0019)}$&$0.0134_{(0.0011)}$&$0.0196_{(0.0031)}$&$0.0118_{(0.0016)}$\\
$p=300$&GFP&$0.0109_{(0.0051)}$&$0.0224_{(0.0172)}$&$0.1391_{(0.1021)}$&$0.0138_{(0.0127)}$&$0.1327_{(0.0413)}$&$0.1950_{(0.0341)}$\\
&GFN&$0.0023_{(0.0026)}$&$0_{(0)}$&$0_{(0)}$&$0_{(0)}$&$0_{(0)}$&$0.0035_{(0.0025)}$\\
\midrule
&MSE&$0.0012_{(0.0006)}$&$0.0025_{(0.0003)}$&$0.0016_{(0.0006)}$&$0.0023_{(0.0003)}$&$0.0071_{(0.0019)}$&$0.0021_{(0.006)}$\\
$n=300$&MAE&$0.0032_{(0.0007)}$&$0.0046_{(0.0003)}$&$0.0041_{(0.0006)}$&$0.0045_{(0.0003)}$&$0.0100_{(0.0022)}$&$0.0066_{(0.0008)}$\\
$p=900$&GFP&$0.0023_{(0.0022)}$&$0.0132_{(0.0068)}$&$0.0568_{(0.0343)}$&$0.0069_{(0.0049)}$&$0.1074_{(0.0331)}$&$0.1543_{(0.0202)}$\\
&GFN&$0.0014_{(0.0010)}$&$0_{(0)}$&$0_{(0)}$&$0_{(0)}$&$0_{(0.0002)}$&$0.0021_{(0.0007)}$\\
\midrule
&MSE&$0.0010_{(0.0004)}$&$0.0015_{(0.0001)}$&$0.0010_{(0.0004)}$&$0.0014_{(0.0002)}$&$0.0046_{(0.0013)}$&$0.0016_{(0.0003)}$\\
$n=300$&MAE&$0.0022_{(0.0005)}$&$0.0027_{(0.0002)}$&$0.0025_{(0.0004)}$&$0.0027_{(0.0002)}$&$0.0078_{(0.0022)}$&$0.0050_{(0.0005)}$\\
$p=1500$&GFP&$0.0013_{(0.0012)}$&$0.0092_{(0.0044)}$&$0.0405_{(0.0263)}$&$0.0047_{(0.0029)}$&$0.0921_{(0.0261)}$&$0.1274_{(0.0160)}$\\
&GFN&$0.0011_{(0.0006)}$&$0_{(0)}$&$0_{(0)}$&$0_{(0)}$&$0.0001_{(0.0002)}$&$0.0018_{(0.0005)}$\\
\bottomrule
    \end{tabular}
  \end{center}
\end{table}

\section{Real Data Analysis}\label{real data}
We apply the proposed method to the birth weight dataset collected at Baystate Medical Center, Springfield, Massachusetts, in 1986. This dataset comprises birth weights of 189 infants and eight maternal predictors. We use the preprocessed \texttt{Birthwt} dataset available in R, which contains 189 observations with 16 predictors. For each of 100 random splits, $80\%$ of observations form the training set and the remaining $20\%$ form the test set. We evaluate MAE, MSE, and computational time as defined in Section~\ref{Timing Comparisons}.
\begin{table}[htbp]
  \begin{center}
  \small
    \caption{Results for the \texttt{Birthwt} Dataset.}\label{CC1}
    \begin{tabular}{cccc} 
\toprule
    &SGL-DADMM&HAQ-GMD&GPQR\\
\cmidrule(lr){2-2}\cmidrule(lr){3-3}\cmidrule(lr){4-4}
~&~&$\tau=0.25$&\\
\midrule
MSE&$3.2850_{(0.6486)}$&$9.2767_{(0.6124)}$&$9.2750_{(0.6023)}$\\
MAE&$1.4529_{(0.1694)}$&$2.9599_{(0.1051)}$&$2.9616_{(0.1039)}$\\
time&$0.01_{(0.01)}$&$0.03_{(0.01)}$&$0.04_{(0.01)}$\\
\midrule
~&~&$\tau=0.5$&\\
\midrule
MSE&$3.0163_{(0.5844)}$&$9.3801_{(0.6497)}$&$9.2957_{(0.6012)}$\\
MAE&$1.3881_{(0.1525)}$&$2.9781_{(0.1118)}$&$2.9677_{(0.1036)}$\\
time&$0.01_{(0.01)}$&$0.03_{(0.01)}$&$0.03_{(0.01)}$\\
\midrule
~&~&$\tau=0.75$&\\
\midrule
MSE&$3.6669_{(0.6299)}$&$9.1738_{(0.6318)}$&$9.2078_{(0.6178)}$\\
MAE&$1.5745_{(0.1584)}$&$2.9430_{(0.1093)}$&$2.9523_{(0.1067)}$\\
time&$0.02_{(0.01)}$&$0.04_{(0.01)}$&$0.04_{(0.02)}$\\
\bottomrule
    \end{tabular}
  \end{center}
\end{table}

Table~\ref{CC1} summarizes the performance of SGL-DADMM, HAQ-GMD, and GPQR on the \texttt{Birthwt} dataset. All three methods are computationally fast on this small dataset. SGL-DADMM has a shorter runtime than both competitors and consistently achieves the lowest MSE and MAE across all quantile levels. Although the standard deviations of SGL-DADMM are somewhat larger in certain settings, its accuracy advantage is clear. These results confirm the effectiveness of the proposed algorithm in practice.

\section{Conclusion}\label{conclusion}
In practice, predictors often exhibit a natural grouped structure. This paper considers a linear regression model with $g$ groups of explanatory variables and studies quantile regression with the adaptive sparse group lasso penalty, which simultaneously identifies important groups and selects significant individual variables within those groups. We develop the SGL-DADMM algorithm from a dual perspective and establish its global convergence. Through extensive simulation studies and a real data application, we demonstrate the statistical and computational advantages of the proposed algorithm over existing alternatives.


\section*{Disclosure statement}
No potential conflict of interest was reported by the authors.

\section*{Funding}
This work of Jun Fan was supported by the National Natural Science Foundation of
China under Grant ($11801130$, $12271022$) and the Hebei Natural Science
Foundation ($A2019202135$, $A2023202038$).
\bibliographystyle{nameyear}
\bibliography{refs}

\appendix
\section{Proof of Theorem  3.1}\label{A}
According to the reference \cite{2015An}, $({\btheta}^*,{\bxi}^*)$ is the optimal solution to the dual problem \eqref{shoulianduiou2} if and only if there exists a Lagrange multiplier ${\bLambda}^*$ such that
\begin{equation}
\begin{array}{l}\label{c.1}
{\by}-{\bA}^T{\bLambda}^*=0, \bm{B}^T{\bLambda}^*\in\partial g({\bxi}^*), {\bA}{\btheta}^*+\boldsymbol{B}{\bxi}^*=\bm{0},
\end{array}
\end{equation}
where $\partial g$ denotes the subdifferential of $g$, and any ${\bLambda}^*$ satisfying \eqref{c.1} is the optimal solution to its primal problem.

The proof proceeds in three main steps:

1) Firstly, we establish the boundedness of all variables in the sequence $\left\{\left(\btheta^{k},\bxi^{k},\bLambda^{k}\right)\right\}$. Combining the given definition, the iterative framework of the SGL-DADMM algorithm is equivalent to
\begin{equation*}
\begin{cases}
&\btheta^{k+1}=\mathop{\arg\!\min}\limits_{{\btheta}} \langle{\by},\btheta\rangle-\left\langle{\bLambda}^k,{\bA\btheta}\right\rangle+\dfrac{\sigma}{2}{\left\|\bA{\btheta}+\bm{B}\bxi^k\right\|}^{2}_2; \\
&\bxi^{k+1}=\mathop{\arg\!\min}\limits_{{\bxi}} g(\bxi)-\left\langle{\bLambda}^k,{\bm{B}\bxi}\right\rangle+\dfrac{\sigma}{2}{\left\|\bA{\btheta}^{k+1}+\bm{B}\bxi\right\|}^{2}_2;  \\
&\bLambda^{k+1}=\bLambda^{k}-\mu\sigma(\bA\btheta^{k+1}+\bm{B}\bxi^{k+1}).
\end{cases}
\end{equation*}
In this iterative framework, the optimality condition for $\btheta^{k+1}$ implies
\begin{equation*}
\begin{array}{l}
\by-\bA^T\bLambda^k+\sigma \bA^T\bLambda+\sigma \bA^T(\bA\btheta^{k+1}+\bm{B}\bxi^{k})=0,
\end{array}
\end{equation*}
Combining ${\bLambda}^k={\bLambda}^{k+1}+\mu\sigma\left(\bA\btheta^{k+1}+\bm{B}\bxi^{k+1}\right)$ with $\bA\btheta^{k+1}+\bm{B}\bxi^k=\bA\btheta^{k+1}+\bm{B}\bxi^{k+1}+\bm{B}(\bxi^k-\bxi^{k+1})$, we obtain
\begin{equation}\label{c.2}
\begin{array}{l}
{\by}-{\bA}^T\left[{\bLambda}^{k+1}-\sigma(1-\mu)\left(\bA\bxi^{k+1}+\bm{B}\bgamma^{k+1}\right)-\sigma\bm{B}({\bgamma}^{k}-{\bgamma}^{k+1})\right]=0.
\end{array}
\end{equation}
Similarly, the optimality condition for $\bxi^{k+1}$ implies
\begin{equation*}
\begin{array}{l}
0\in\partial g(\bxi^{k+1})-\bm{B}^T\bLambda^k+\sigma \bm{B}^T(\bA\btheta^{k+1}+\bm{B}\bxi^{k+1}),
\end{array}
\end{equation*}
Combining ${\bLambda}^k={\bLambda}^{k+1}+\mu\sigma\left(\bA\btheta^{k+1}+\bm{B}\bxi^{k+1}\right)$, we obtain
\begin{equation}\label{c.3}
\begin{array}{l}
\bm{B}^T\left[{\bLambda}^{k+1}-\sigma(1-\mu)\left(\bA\btheta^{k+1}+\bm{B}\bxi^{k+1}\right)\right]\in\partial g(\bxi^{k+1}).
\end{array}
\end{equation}
For the optimal solution $\left({\btheta}^*,{\bxi}^*,{\bLambda}^*\right)$ and any $k\geq 1$, let ${\btheta}_e^k={\btheta}^k-{\btheta}^*$, ${\bxi}_e^k={\bxi}^k-{\bxi}^*$, ${\bLambda}_e^k={\bLambda}^k-{\bLambda}^*$.
By utilizing the optimality condition of ${\btheta}^*$ in \eqref{c.1} and the optimality condition of $\btheta^{k+1}$ in \eqref{c.2}, we can obtain
\begin{equation*}
\begin{array}{l}
{\bA}^T[{\bLambda}_e^{k+1}-\sigma(1-\mu)\left(\bA\btheta^{k+1}+\bm{B}\bxi^{k+1}\right)-\sigma\bm{B}({\bxi}^{k}-{\bxi}^{k+1})] =\bm{0},
\end{array}
\end{equation*}
which simplifies to
\begin{equation}\label{c.4}
\begin{array}{l}
\langle{\bLambda}_e^{k+1}-\sigma(1-\mu)\left(\bA\btheta^{k+1}+\bm{B}\bxi^{k+1}\right)-\sigma\bm{B}({\bxi}^{k}-{\bxi}^{k+1}),{\bA}{\btheta}_e^{k+1}\rangle=0.
\end{array}
\end{equation}
Similarly, it can be see that the optimality conditions \eqref{c.1} of ${\bxi}^*$ and the optimality conditions \eqref{c.3} of ${\bxi}^{k+1}$  that
\begin{equation}\label{c.5}
\begin{array}{l}
\langle{\bLambda}_e^{k+1}-\sigma(1-\mu)\left(\bA\btheta^{k+1}+\bm{B}\bxi^{k+1}\right),\bm{B}{\bxi}_e^{k+1}\rangle\geq \bm{0}.
\end{array}
\end{equation}
By combining \eqref{c.4} with \eqref{c.5} and
\begin{equation*}
\begin{array}{l}
\bA{\btheta}_e^{k+1}+\bm{B}{\bxi}_e^{k+1}
=\bA\btheta^{k+1}+\bm{B}\bxi^{k+1}
=(\mu\sigma)^{-1}({\bLambda}^{k}-{\bLambda}^{k+1})=(\mu\sigma)^{-1}({\bLambda}_e^{k}-{\bLambda}_e^{k+1}),
\end{array}
\end{equation*}
we have
\begin{equation}
\begin{array}{l}\label{c.6}
(\sigma\mu)^{-1}\langle{\bLambda}_e^{k+1},{\bLambda}_e^{k}-{\bLambda}_e^{k+1}\rangle-\sigma(1-\mu)\|\bA\btheta^{k+1}+\bm{B}\bxi^{k+1}\|_2^2\\
+\sigma\langle\bm{B}({\bxi}^{k+1}-{\bxi}^k),\bA\btheta^{k+1}+\bm{B}\xi^{k+1}\rangle-\sigma\langle\bm{B}({\bxi}^{k+1}-{\bxi}^k),\bm{B}{\bxi}_e^{k+1}\rangle\geq\bm{0}.
\end{array}
\end{equation}
Note that
$\bm{B}^T\left[{\bLambda}^{k+1}+\sigma(\mu-1)\left(\bA\btheta^{k+1}+\bm{B}\bxi^{k+1}\right)\right]\in\partial g({\bxi}^{k+1})$ and $\bm{B}^T{\bLambda}^{k}+\sigma(\mu-1)\bm{B}^T\\ \left(\bA\btheta^{k}+\bm{B}\bxi^{k}\right)\in\partial g({\bxi}^{k})$.
By optimality of ${\bLambda}^{k+1}={\bLambda}^k-\mu\sigma\left(\bA\btheta^{k+1}+\bm{B}\bxi^{k+1}\right)$, we obtain that
\begin{equation*}
\begin{split}
0&\leq\big\langle{\bLambda}^{k+1}-{\bLambda}^k+\mu\sigma\left(\bA\btheta^{k+1}+\bm{B}\bxi^{k+1}\right)-\sigma\left(\bA\bxi^{k+1}+\bm{B}\bgamma^{k+1}\right)\\
&~~~-\sigma(\mu-1)\left(\bA\btheta^{k}+\bm{B}\bxi^{k}\right),\bm{B}\left({\bxi}^{k+1}-{\bxi}^{k}\right)\big\rangle,
\end{split}
\end{equation*}
which implies
\begin{equation*}
\begin{array}{l}
\sigma\big\langle \bA\btheta^{k+1}+\bm{B}\bxi^{k+1},\bm{B}({\bxi}^{k+1}-{\bxi}^{k})\big\rangle
\leq\sigma(1-\mu)\big\langle \bA\btheta^{k}+\bm{B}\bxi^{k},\bm{B}({\bxi}^{k+1}-{\bxi}^{k})\big\rangle.
\end{array}
\end{equation*}
The formula \eqref{c.6} becomes
\begin{equation*}
\begin{array}{l}
(\mu\sigma)^{-1}\langle{\bLambda}_e^{k+1},{\bLambda}_e^{k}-{\bLambda}_e^{k+1}\rangle-\sigma(1-\mu)\|\bA\btheta^{k+1}+\bm{B}\bxi^{k+1}\|_2^2\\
+\sigma(1-\mu)\langle\bm{B}({\bxi}^{k+1}-{\bxi}^{k}),\bA\btheta^{k}+\bm{B}\bgamma^{k}\rangle
-\sigma\langle\bm{B}({\bxi}^{k+1}-{\bxi}^{k}),\bm{B}{\bxi}_e^{k+1}\rangle\geq 0.
\end{array}
\end{equation*}

Applying  $2\langle{\bu},{\bv}\rangle=\|{\bu}\|_2^2+\|{\bv}\|_2^2-\|{\bu}-{\bv}\|_2^2$ and ${\bLambda}_e^{k+1}-{\bLambda}_e^{k}={\bLambda}^{k+1}-{\bLambda}^{k}=-\mu\sigma(\bA\btheta^{k+1}+\bm{B}\bxi^{k+1})$, we further obtain that
\begin{equation}
\begin{array}{l}\label{c.7}
(\mu\sigma)^{-1}(\|{\bLambda}_e^{k+1}\|_2^2-\|{\bLambda}_e^{k}\|_2^2)-\sigma(2-\mu)\|\bA\btheta^{k+1}+\bm{B}\bxi^{k+1}\|_2^2\\
+2\sigma(1-\mu)\langle\bm{B}({\bxi}^{k+1}-{\bxi}^{k}),\bA\btheta^{k}+\bm{B}\bxi^{k}\rangle
-\sigma\|\bm{B}({\bxi}^{k+1}-{\bxi}^{k})\|_2^2\\
-\sigma\|\bm{B}{\bxi}_e^{k+1}\|_2^2+\sigma\|\bm{B}{\bxi}_e^{k}\|_2^2\geq 0.
\end{array}
\end{equation}
The following considers two cases: $0 <\mu\leq 1$ and $1<\mu<(1+\sqrt{5})/2$. We will prove that the sequence generated by the SGL-DADMM algorithm converges when $\mu\in(0, (1+\sqrt{5})/2)$.
\begin{enumerate}
\item[(i)]$0<\mu\leq1$, note that $\min\{\mu,1+\mu-\mu^2\}=\mu$ and $\min\{\mu,\mu^{-1}\}=\mu$, by Cauchy-Schwarz inequality we have
\begin{equation*}
\begin{array}{l}
2\sigma(1-\mu)\langle\bm{B}({\bxi}^{k+1}-{\bxi}^{k}),\bA\btheta^{k}+\bm{B}\bxi^{k}\rangle
\\ \leq\sigma(1-\mu)[\|\bm{B}({\bxi}^{k+1}-{\bxi}^{k})\|_2^2
+\|\bA\btheta^{k}+\bm{B}\bxi^{k}\|_2^2] .
\end{array}
\end{equation*}
It follows from \eqref{c.7} that,
\begin{equation}
\begin{array}{l}\label{cc}
\left[(\mu\sigma)^{-1}\|{\bLambda}_e^{k+1}\|_2^2+\sigma\|\bm{B}{\bxi}_e^{k+1}\|_2^2+\sigma(1-\mu)\|\bA\btheta^{k+1}+\bm{B}\bxi^{k+1}\|_2^2\right]\\
-\left[(\mu\sigma)^{-1}\|{\bLambda}_e^{k}\|_2^2+\sigma\|\bm{B}{\bxi}_e^{k}\|_2^2+\sigma(1-\mu)\|\bA\btheta^{k}+\bm{B}\bxi^{k}\|_2^2\right]\\
\leq-\sigma\|\bA\btheta^{k+1}+\bm{B}\bxi^{k+1}\|_2^2-\sigma\mu\|\bm{B}({\bxi}^{k+1}-{\bxi}^{k})\|_2^2.
\end{array}
\end{equation}
For notational convenience, denote
\begin{equation*}
\begin{array}{l}
{\ba}_{k+1}=(\mu\sigma)^{-1}\|{\bLambda}_e^{k+1}\|_2^2+\sigma\|\bm{B}{\bxi}_e^{k+1}\|_2^2,\\
{\bm{b}}_{k+1}=\min\{\mu,1+\mu-\mu^2\}\sigma\|\bm{B}({\bgamma}^{k+1}-{\bgamma}^{k})\|_2^2,
\end{array}
\end{equation*}
We obtain from \eqref{cc} that
\begin{equation}
\begin{array}{l}\label{c.8}
\left[{\ba}_{k+1}+\sigma(1-\mu)\|\bA\btheta^{k+1}+\bm{B}\bxi^{k+1}\|_2^2\right]-\left[{\ba}_{k}+\sigma(1-\mu)\|\bA\btheta^{k}+\bm{B}\bxi^{k}\|_2^2\right]\\
\leq-\bm{b}_{k+1}-\sigma\|\bA\btheta^{k+1}+\bm{B}\bxi^{k+1}\|_2^2.
\end{array}
\end{equation}
\item[(ii)]$1<\mu<(1+\sqrt{5})/2$, note that $\min\{\mu,1+\mu-\mu^2\}=1+\mu-\mu^2$  and $\min\{\mu,\mu^{-1}\}=\mu^{-1}$, by Cauchy-Schwarz inequality we have:
\begin{equation*}
\begin{array}{l}
2\sigma(1-\mu)\langle\bm{B}({\bxi}^{k+1}-{\bxi}^{k}),\bA\btheta^{k}+\bm{B}\bxi^{k}\rangle
\\ \leq\sigma(\mu-1)[\mu\|\bm{B}({\bxi}^{k+1}-{\bxi}^{k})\|_2^2
+\mu^{-1}\|\bA\btheta^{k}+\bm{B}\bxi^{k}\|_2^2] .
\end{array}
\end{equation*}
Using equation \eqref{c.7} again, we can deduce
\begin{equation}
\begin{array}{l}\label{c.9}
\left[{\ba}_{k+1}+\sigma(1-\mu^{-1})\|\bA\btheta^{k+1}+\bm{B}\bxi^{k+1}\|_2^2\right]
-\left[{\ba}_{k}+\sigma(1-\mu^{-1})\|\bA\btheta^{k}+\bm{B}\bxi^{k}\|_2^2\right]\\
\leq-\bm{b}_{k+1}-\sigma(1-\mu+\mu^{-1})\|\bA\btheta^{k+1}+\bm{B}\bxi^{k+1}\|_2^2.
\end{array}
\end{equation}
In conclusion, for all $\mu \in (0, (1+\sqrt{5})/2)$, equations \eqref{c.8} and \eqref{c.9} can be uniformly expressed as
\begin{equation}
\begin{array}{l}\label{c.10}
\left[{\ba}_{k+1}+(1-\min\{\mu,\mu^{-1}\})\sigma\|\bA\btheta^{k+1}+\bm{B}\bxi^{k+1}\|_2^2\right]\\
-\left[{\ba}_{k}+(1-\min\{\mu,\mu^{-1}\})\sigma\|\bA\btheta^{k}+\bm{B}\bxi^{k}\|_2^2\right]\\
\leq-\left[\bm{b}_{k+1}+(1-\mu+\min\{\mu,\mu^{-1}\})\sigma\|\bA\btheta^{k+1}+\bm{B}\bxi^{k+1}\|_2^2\right].
\end{array}
\end{equation}
\end{enumerate}
Let $\bc_k={\ba}_{k}+(1-\min\{\mu,\mu^{-1}\})\sigma\|\bA\btheta^{k}+\bm{B}\bxi^{k}\|_2^2$, equation \eqref{c.10} indicates that $\bc_k$ is a non-increasing sequence bounded below by 0 and bounded above by $\bc_0 < \infty$, thus the sequence $\{\bc_k\}$ converges. According to the definition of ${\ba}_{k+1}$, it follows that $\{\|{\bLambda}^{k+1}\|_2\}$ and $\{\|{\bxi}_e^{k+1}\|_{\bm{B}^T\bm{B}}^2\}$ are bounded sequences. Since $\bm{B}^T\bm{B}$ is a positive definite matrix, the sequence $\{\|{\bxi}^{k}\|_2\}$ is also bounded. Moreover, we have
\begin{equation*}
\begin{array}{l}
0\leq\lim\limits_{t\to\infty}\bc_t=\bc_0+\lim\limits_{t\to\infty}\sum_{k=0}^{t-1}(\bc_{k+1}-\bc_k),
\end{array}
\end{equation*}
thus$-\lim\limits_{t\to\infty}\bc_t=-\bc_0+\lim\limits_{t\to\infty}\sum_{k=0}^{t-1}|\bc_{k+1}-\bc_k|$. Therefore, the sequence $\{\bc_{k+1}-\bc_k\}$ converges absolutely, $\lim\limits_{t\to\infty}|\bc_{k+1}-\bc_k|=0$. Combining $|\bc_{k+1}-\bc_k|\geq \bm{b}_{k+1}+(1+\mu+\min\{\mu,\mu^{-1}\})\sigma\|\bA\btheta^{k+1}+\bm{B}\bxi^{k+1}\|_2^2$, we have
\begin{equation}\label{c.11}
\begin{split}
&\lim\limits_{k\to\infty}\bm{b}_{k+1}=0,\lim\limits_{k\to\infty}\|{\bLambda}^{k+1}-{\bLambda}^k\|_2=\lim\limits_{k\to\infty}\mu\sigma\|\bA\btheta^{k+1}+\bm{B}\bxi^{k+1}\|_2^2=0\text{.}
\end{split}
\end{equation}
Moreover, it is noted that
\begin{equation}\label{c.12}
\begin{split}
\|{\bA}{\btheta}_{e}^{k+1}\|_2&\leq\|{\bA}{\btheta}_{e}^{k+1}+\bm{B}{\bxi}_{e}^{k+1}\|_2+\|\bm{B}{\bxi}_{e}^{k+1}\|_2\\
&=\|\bA\btheta^{k+1}+\bm{B}\bxi^{k+1}\|_2+\|\bm{B}{\bxi}_{e}^{k+1}\|_2\text{.}
\end{split}
\end{equation}
We can obtain that the sequence $\{\|{\bA}{\btheta}_{e}^{k+1}\|_2\}$ is bounded. Therefore, $\{\|{\btheta}_{e}^{k+1}\|_{{\bA}^T{\bA}}^2\}$ is bounded, and since ${\bA}^T{\bA}$ is positive definite, it implies that the sequence $\{\|{\theta}^{k+1}\|_2\}$ is bounded. In summary, the iterative sequence $\left\{({\btheta}^k,{\bxi}^k,{\bLambda}^k),k\geq1\right\}$ is bounded, and we can find a subsequence $\left\{({\btheta}^{k_m},{\bxi}^{k_m},{\bLambda}^{k_m}),m\geq1\right\}$ that converges to some limit point $\left({\btheta}^{\infty},{\bxi}^{\infty},{\bLambda}^{\infty}\right)$.

~2)The following proves that $\left(\btheta^{\infty},\bxi^{\infty}\right)$ is the optimal solution to problem \eqref{shoulianduiou2}, with $\bLambda^{\infty}$ being the corresponding Lagrange multiplier. First, by \eqref{c.11}, we have
\begin{equation}\label{c.13}
\begin{array}{l}
\lim\limits_{k\to\infty}\|\bm{B}({\bxi}^{k+1}-{\bxi}^{k})\|_2=0\text{.}
\end{array}
\end{equation}
Using the inequality
\begin{equation*}
\begin{array}{l}
\|{\bA}{\btheta}^{k+1}+\bm{B}{\bxi}^{k}\|_2\leq\|{\bA}{\btheta}^{k+1}+\bm{B}{\bxi}^{k+1}\|_2+\|\bm{B}({\bxi}^{k+1}-{\bxi}^{k})\|_2,
\end{array}
\end{equation*}
we deduce from \eqref{c.11} and \eqref{c.13} that
\begin{equation}\label{c.14}
\begin{array}{l}
\lim\limits_{k\to\infty}\|{\bLambda}^{k+1}-{\bLambda}^{k}\|_2=0, \lim\limits_{k\to\infty}\|{\bA}{\bxi}^{k+1}+\bm{B}{\bgamma}^{k}\|_2=0\text{.}
\end{array}
\end{equation}
Furthermore, by the convexity of $\delta_{\mathcal{C}}$, it is known that $\partial g$ is a non-empty closed set. Taking the limit as $k \to \infty$ on both sides of equations \eqref{c.2} and \eqref{c.3}, we obtain
\begin{equation*}
\begin{array}{l}
{\by}-{\bA}^T{\bLambda}^{\infty}=0, \bm{B}^T{\bLambda}^{\infty}\in\partial g({\bxi}^{\infty}), {\bA}{\btheta}^{\infty}+\bm{B}{\bxi}^{\infty}={0},
\end{array}
\end{equation*}
thus $\left({\btheta}^{\infty},{\bxi}^{\infty},{\bLambda}^{\infty}\right)$ satisfies equation \eqref{c.1}. Combining this with the earlier discussion of equation \eqref{c.1}, it can be concluded that $({\btheta}^{\infty},{\bxi}^{\infty})$ is the optimal solution to the dual problem \eqref{shoulianduiou2}, and $\bLambda^{\infty}$ is the corresponding Lagrange multiplier.

~3)Finally, it is proven that $({\btheta}^{\infty},{\bxi}^{\infty},{\bLambda}^{\infty})$ is the unique accumulation point of the sequence $\{({\btheta}^k,{\bxi}^k,{\bLambda}^k)\}$. Since $({\btheta}^{\infty},{\bxi}^{\infty},{\bLambda}^{\infty})$ satisfies \eqref{c.1}, it can be replaced by $({\btheta}^{*},{\bxi}^{*},{\bLambda}^{*})$. From \eqref{c.11}, as $m \to \infty$, the subsequence $\{{\ba}_{k_m}+(1-\min\{\mu,\mu^{-1}\})\sigma\|\bA{\btheta}^{k_m}+\bm{B}{\xi}^{k_m}\|_2^2\}$ converges to 0. Since this subsequence comes from a non-increasing sequence and the sequence $\{\|\bA\btheta^{k}+\bm{B}\bxi^{k}\|_2^2\}$ converges to 0, we have $\lim\limits_{k\to\infty}{\ba}^k={0}$. Therefore, we have $\lim\limits_{k\to\infty}{\bLambda}^k={\bLambda}^{\infty}$, and combined with \eqref{c.12}, we obtain $\lim\limits_{k\to\infty}\|\bm{B}{\bxi}_e^k\|_2^2={0}$. Since $\bm{B}^T\bm{B}$ is a positive definite matrix, we have $\lim\limits_{k\to\infty}{\bxi}^k={\bxi}^{\infty}$. Equation \eqref{c.12} implies $\lim\limits_{k\to\infty}\|{\bA}{\btheta}_e^k\|_2={0}$. Combining this with the positive definiteness of the matrix ${\bA}^T{\bA}$, we can deduce $\lim\limits_{k\to\infty}{\btheta}^k={\btheta}^{\infty}$.

In conclusion, if $\mu \in \left(0,(1+\sqrt{5})/2\right)$, then $\lim\limits_{k\to\infty}\left({\btheta}^k,{\bxi}^k,{\bLambda}^k\right)=({\btheta}^{\infty},{\bxi}^{\infty},{\bLambda}^{\infty})$.

\end{document}